\documentclass[sn-mathphys-num]{sn-jnl}

\usepackage[T1]{fontenc}
\usepackage{lmodern}

\usepackage{graphicx}%
\usepackage{multirow}%
\usepackage{amsmath,amssymb,amsfonts}%
\usepackage{amsthm}%
\usepackage[title]{appendix}%
\usepackage{xcolor}%
\usepackage{textcomp}%
\usepackage{manyfoot}%
\usepackage{booktabs}%
\usepackage{algorithm}%
\usepackage{algorithmicx}%
\usepackage{algpseudocode}%
\usepackage{listings}%

\usepackage{amsthm}

\newtheoremstyle{thmstyleone}
  {3pt} 
  {3pt} 
  {\itshape} 
  {} 
  {\bfseries} 
  {.} 
  { } 
  {} 

\newtheoremstyle{thmstyletwo}
  {3pt}
  {3pt}
  {}
  {}
  {\bfseries}
  {.}
  { }
  {}

\newtheoremstyle{thmstylethree}
  {3pt}
  {3pt}
  {}
  {}
  {\itshape}
  {.}
  { }
  {}
  
\theoremstyle{plain} 
\theoremstyle{thmstyletwo}%

\theoremstyle{thmstylethree}%
\usepackage{xcolor}%
\begin{document}

\title[Grand Challenges in Immersive Technologies for Cultural Heritage]{Grand Challenges in Immersive Technologies for Cultural Heritage}


\author[1]{\fnm{Hanbing} \sur{Wang}}\email{wanghanbing@mail.tsinghua.edu.cn} 

\author[1]{\fnm{Junyan} \sur{Du}}\email{djy23@mails.tsinghua.edu.cn} 

\author[2]{\fnm{Yue} \sur{Li}}\email{yue.li@xjtlu.edu.cn} 

\author[1]{\fnm{Lie} \sur{Zhang}}\email{zhlie@tsinghua.edu.cn} 

\author*[3,4]{\fnm{Xiang} \sur{Li}}\email{xl529@cam.ac.uk} 

\affil[1]{\orgdiv{Institute of Interactive Media, Academy of Fine Arts}, \orgname{Tsinghua University}, \orgaddress{\city{Beijing}, \country{China}}}

\affil[2]{\orgdiv{Department of Computing}, \orgname{Xi'an Jiaotong-Liverpool University}, \orgaddress{\city{Suzhou}, \country{China}}}

\affil[3]{\orgdiv{Department of Engineering}, \orgname{University of Cambridge}, \orgaddress{\city{Cambridge}, \country{United Kingdom}}}

\affil[4]{\orgdiv{Leverhulme Centre for the Future of Intelligence}, \orgname{University of Cambridge}, \orgaddress{\city{Cambridge}, \country{United Kingdom}}}


\abstract{Cultural heritage, a testament to human history and civilization, has gained increasing recognition for its significance in preservation and dissemination. The integration of immersive technologies has transformed how cultural heritage is presented, enabling audiences to engage with it in more vivid, intuitive, and interactive ways. However, the adoption of these technologies also brings a range of challenges and potential risks. This paper presents a systematic review, with an in-depth analysis of 177 selected papers. We comprehensively examine and categorize current applications, technological approaches, and user devices in immersive cultural heritage presentations, while also highlighting the associated risks and challenges. Furthermore, we identify areas for future research in the immersive presentation of cultural heritage. Our goal is to provide a comprehensive reference for researchers and practitioners, enhancing understanding of the technological applications, risks, and challenges in this field, and encouraging further innovation and development.
}




\maketitle

\section{Introduction}\label{sec1}

The United Nations Educational, Scientific and Cultural Organization (UNESCO) aims to promote the identification, protection, and preservation of cultural and natural heritage worldwide, recognizing sites of outstanding value to humanity\footnote{\href{https://whc.unesco.org/en/conventiontext}{https://whc.unesco.org/en/conventiontext}}. Nowadays, the use of immersive technologies in cultural heritage displays has increasingly become the focus of academic research, as institutions seek to enhance visitor engagement and broaden access to valuable artifacts and experiences~\citep{bekele_survey_2018}. These immersive displays combine technologies such as virtual reality (VR), augmented reality (AR), mixed reality (MR), and other platforms like 360-degree video and interactive digital installations, enabling audiences to experience cultural heritage —here encompassing both tangible and intangible forms— in more vivid, intuitive, and personal ways. However, the adoption of these technologies also introduces several risks and challenges.

In the realm of immersive technologies for cultural heritage, prior studies have explored a variety of issues. Some research emphasizes digital preservation and conservation~\citep{evens_challenges_2011}, while others focus on specific applications of immersive technologies or isolated aspects of cultural heritage~\citep{boboc_augmented_2022, chong_virtual_2021, buragohain_digitalizing_2024, idris_preservation_2016, koller2010research, beraldin2005virtual, windhager_visualization_2019, baratin_what_2023, innocente2023framework, wei2024meta}. However, certain works neglect to address the challenges or negative impacts—often referred to as the ``dark sides''—associated with these technologies~\citep{zhang_reviving_2024, bekele_comparison_2019, wagner_safeguarding_2023, nofal_eslam_phygital_nodate, kukreja_digital_2024}. Additionally, in many cases, the application of immersive technologies to cultural heritage is restricted to VR, AR, and MR~\citep{zhang2024impact}. Furthermore, some studies rely on samples that predate 2018~\citep{bekele_survey_2018, piccialli2017cultural}, or their sample sizes are too small to draw broader conclusions~\citep{piccialli2017cultural}.

Given these gaps, it is necessary to conduct a new survey of the literature on immersive technologies for cultural heritage. This review will place particular emphasis on the grand challenges of applying immersive technologies to digital cultural heritage, offering a critical reflection on their impact in this field.

In this paper, we set the stage for a comprehensive review of the literature on immersive technologies in cultural heritage. Our review covers a pool of 5,368 articles published up to January 27, 2024, sourced from leading databases including the ACM Digital Library, IEEE Xplore, and Scopus. From this, we systematically analyzed 177 papers to identify these papers according to three main categories: ``device,'' ``application,'' and ``technology.'' Based on our analysis, we identified and summarized the strengths of the equipment, applications, and technologies, which can be grouped into the following areas: enhancing participation and experience, fostering social engagement, advancing education and knowledge dissemination, optimizing technological implementation, and promoting heritage protection and transmission.

We also identified and categorized the shortcomings of the equipment, applications, and technologies, focusing on aspects such as technical barriers, sustainability, user experience, and safety concerns. Additionally, we found certain areas that have not been fully explored in the research, which could have a significant impact on the digital display and dissemination of cultural heritage. These gaps include the potential misinterpretation of cultural heritage, the influence on media purity, and the possible potential damage to cultural heritage.

By exploring these diverse dimensions, our goal is to offer a comprehensive reference that will inform both researchers and practitioners, supporting more informed decision-making and fostering innovative solutions in the field of cultural heritage preservation in the digital age.

\section{Background}
\subsection{Research Status of Cultural Heritage Immersion Technology}

Immersive technology blurs the boundary between the physical and virtual worlds, allowing users to experience a heightened sense of immersion~\citep{lee2013system}. This category includes AR, VR, MR, as well as haptic technology, remote immersion, and more~\citep{suh2018state}. Immersive technology captures an individual’s physical movements, posture, and gestures as inputs, which are then used to interact with the virtual environment, making the user feel as if they are truly present in that environment~\citep{handa2012immersive}. When applied to the display of cultural heritage, immersive technologies enhance the quality and dissemination of cultural content by stimulating our senses in a more vivid and natural way~\citep{innocente2023framework}.

To understand the current state of immersive technology research in the context of cultural heritage, we conducted two searches on Google Scholar. The first search used the keywords “\textsc{cultural heritage}” and “\textsc{immersive technology},” while the second search added “\textsc{challenge}” to refine the focus. From these searches, we selected relevant literature review articles. By analyzing these articles, we aimed to develop a comprehensive understanding of the research landscape. However, these review articles present several limitations:



\begin{itemize}

\item They primarily focus on issues related to digital preservation and protection, rather than the presentation aspect of cultural heritage~\citep{evens_challenges_2011}.

\item They discuss only specific applications of immersive technology or focus on isolated aspects of cultural heritage, without providing a broader perspective~\citep{boboc_augmented_2022, chong_virtual_2021, buragohain_digitalizing_2024, idris_preservation_2016, koller2010research, beraldin2005virtual, windhager_visualization_2019, baratin_what_2023, innocente2023framework}.

\item They do not address the challenges or negative implications of the technologies~\citep{zhang_reviving_2024, bekele_comparison_2019, wagner_safeguarding_2023, nofal_eslam_phygital_nodate, kukreja_digital_2024}.

\item The immersive technologies applied to cultural heritage are restricted to VR, AR, and MR~\citep{zhang2024impact}.

\item Much of the literature was published before 2018~\citep{bekele_survey_2018, piccialli2017cultural}, or the sample sizes in some studies were too small~\citep{piccialli2017cultural}.

\end{itemize}

Evens and Hauttekeete identify four major issues hindering the sustainability of digital preservation in cultural heritage institutions~\citep{evens_challenges_2011}. Boboc and Chong focus on the application of AR to cultural heritage, with Boboc identifying eight key topics~\citep{boboc_augmented_2022, chong_virtual_2021}, while Buragohain explores the challenges and strategies of digitization in the metaverse~\citep{buragohain_digitalizing_2024}. Idris discusses the technical challenges of protecting intangible cultural heritage~\citep{idris_preservation_2016}, and Koller examines the research challenges in digital archives for 3D cultural relics~\citep{koller2010research}. Beraldin highlights the opportunities and challenges introduced by recent 3D technologies for heritage conservation and virtual reconstruction~\citep{beraldin2005virtual}, while Windhager tackles the difficulties in visualizing collected data~\citep{windhager_visualization_2019}, and Baratin examines issues related to disseminating technical content~\citep{baratin_what_2023}.

Several studies discuss the use of advanced digital technologies in cultural heritage but overlook challenges or negative aspects. Zhang promotes the potential of digital technologies without addressing limitations~\citep{zhang_reviving_2024}, Bekele compares immersive reality technologies without exploring challenges~\citep{bekele_comparison_2019}, and Wagner addresses legal issues but not practical challenges~\citep{wagner_safeguarding_2023}. Nofal and Kukreja delve into ``digital heritage'' and its applications but avoid discussing the ``dark side'' of these technologies~\citep{nofal_eslam_phygital_nodate, kukreja_digital_2024}, while Innocente identifies limitations in the field without covering negative implications~\citep{innocente2023framework}.

Zhang provides a bibliometric analysis of VR, AR, and MR technologies applied to cultural heritage but limits the discussion to these technologies~\citep{zhang2024impact}. Bekele offers an analysis of immersive technologies in cultural heritage from before 2018, noting potential barriers but missing recent advancements~\citep{bekele_survey_2018}. Piccialli summarizes 12 papers published in 2016, offering an overview rather than in-depth analysis~\citep{piccialli2017cultural}.

\subsection{Application of Immersive Technology in Cultural Heritage}

The application of immersive technologies such as VR, AR, MR, and 360-degree video experiences has become an increasingly important area of research in cultural heritage, with a focus on enhancing visitor engagement and expanding access to artifacts and experiences. This section reviews key literature, covering the use of immersive technology in environmental reconstruction, data protection, visitor experience, education, and storytelling.

Kersten et al.~\citep{kersten2018virtual} highlight the role of immersive technology in environmental reconstruction, outlining a two-part workflow: 3D data recording and VR system development, which includes motion navigation, interaction, and multi-user design. Their work compares user experiences in virtual and real-world environments, showcasing how immersive technology connects the virtual with the physical world~\citep{kersten2018virtual}. Similarly, Hassani emphasizes the role of digital technologies in speeding up data collection, recording, and ensuring precise outputs for heritage documentation~\citep{hassani2015documentation}.

For virtual tours of fragile or hard-to-reach historical sites, Ruffin and Permadi explore how immersive technologies, such as VR models, enhance inclusivity by overcoming access barriers for people with disabilities and the elderly~\citep{ruffino2019digital}. Their attention to audience participation, including the use of binaural audio for realistic soundscapes, offers a more humanistic approach to virtual heritage experiences~\citep{ruffino2019digital}.

On a larger scale, Argyriou et al. combine 360-degree immersive video with gamification and narrative-driven virtual experiences to create interactive urban environments~\citep{argyriou_design_2020}. They argue that combining high participation through narrative and freedom through movement technology leads to deeper, long-lasting audience engagement~\citep{argyriou_design_2020}.

In cultural heritage education, Anastasovitis and Roumeliotis advocate for immersive technology to provide safe, controlled environments for both learners and heritage sites~\citep{anastasovitis2020creative}. VR is used, for example, to train underwater archaeologists in excavation tasks while protecting fragile artifacts~\citep{anastasovitis2020creative}. These technologies also allow for remote learning and virtual visits, proving valuable in situations where physical access is restricted~\citep{anastasovitis2020creative}.

Overall, immersive technologies are proving successful in enhancing visitor experience and education in cultural heritage, though the literature suggests varying levels of adoption across sectors, pointing to the need for more standardized reviews and practices.

\subsection{Challenges and Limitations}

While immersive technologies offer significant potential for the digital display of cultural heritage, they present several challenges and limitations.

One major issue is the digital divide and unequal access to advanced technology. Hassani notes that cultural heritage professionals must continually update their skills to manage digital projects effectively~\citep{hassani2015documentation}. However, knowledge gaps between heritage experts, IT specialists, and surveyors, as well as disparities across different regions and socio-economic groups, exacerbate the digital divide~\citep{hassani2015documentation}. Storeide et al. highlighted data heterogeneity and interoperability challenges~\citep{storeide2023standardization}. Differences in how virtual and real-world research emphasize either observer perception or quantitative parameters can lead to issues with the authenticity and integrity of digital representations~\citep{storeide2023standardization}. As a result, it is often difficult to reuse data across different applications due to incompatible formats, and current standards fail to ensure uniform measurements and workflows~\citep{storeide2023standardization}.

User experience and technology adaptation also pose challenges. Argyriou et al. emphasized the need to improve user comfort by addressing factors such as high-resolution video, smooth navigation, and guided interfaces~\citep{argyriou_design_2020}. However, too much guidance can reduce immersion, and finding the right balance between comfort and immersion remains an ongoing challenge.

Regarding cultural appropriation and the commercial use of digital heritage have also arisen. Klinowski and Szafarowicz discuss conflicts between copyright, ownership, and public versus commercial interests in the digitization of cultural heritage~\citep{klinowski2023digitisation}. This illustrates that balancing the rights of creators and owners with public access and commercial use could still be a major challenge. However, the potential of digital approaches to promote wider access and create business opportunities still requires validation through further empirical research.

These challenges highlight the need for a careful, balanced approach when adopting immersive technologies in cultural heritage, ensuring that technological progress supports rather than undermines cultural preservation.

\subsection{Research Gaps and Emergent Strategies}

Despite the growing literature on the use of immersive technologies in various fields, there are still significant gaps in comprehensive, systematic reviews that focus specifically on their use in cultural heritage. Previous surveys have often touched on the potential and benefits of these technologies, but have largely overlooked a systematic exploration of the associated challenges and negative aspects, particularly the unintended consequences of digitization on cultural heritage. These include, but are not limited to, issues such as digital dependence and resistance, social ethics in the digital context, and the misinterpretation of cultural heritage caused by its diverse digital representations. Another concern is the impact of digital technology on the ``purity of media experience'' in presenting cultural heritage. This concept refers to the authenticity and immediacy with which audiences engage with cultural heritage. Ideally, this experience should reflect the original historical and sensory context, without significant mediation or alteration. When individuals engage with cultural heritage, the goal is for them to encounter it in its unaltered form, minimally influenced by interpretive layers, technological enhancements, or information filtering. This principle emphasizes that the processes of media translation and interpretation should not substantially distort the original essence of cultural heritage. Further, the use of digital display technologies in cultural heritage can have potentially damaging effects, including physical damage, light and noise pollution, historical and cultural distortions, and threats to privacy and data security.

Existing reviews of the relationship between immersive technology, cultural heritage and people have often focused on specific aspects of research on how technology helps people to enhance their understanding of cultural heritage, but in our view this is not a comprehensive definition of the relationship between the three. For example, Bentley et al. describe the relationship between digital technology, the field of artificial intelligence, and people~\citep{bentley2024digital}. They identify social factors, as well as digital engagement factors under the influence of digital technologies and digital perception and experience factors as digital confidence that can influence people's attitudes toward AI~\citep{bentley2024digital}. Throughout the paper, the definition of the relationship between the three is expressed as a pattern of interdependence, which is often overlooked in the study of immersive technology and cultural heritage~\citep{bentley2024digital}. Similarly, future research should focus on digital dependence and resistance under the influence of immersive technology, which may have an impact on the understanding of cultural heritage.

Moreover, the social injustices that exist in digital discourse should not be ignored. Johnson focused on the infiltration and substitution of social identity into technology, including the role presetting of gender, race, class, educational background and other factors~\citep{johnson2004technological}. That is, the maximum value of technological progress is defined as being associated with white, middle class, and male, and the technical, scientific, and logical disciplines are also considered to be gender-related. Under such a trend, the development direction of technological progress will be further restricted, and the protection and dissemination of cultural heritage will also fall into a single situation. The research perspective of this paper complements our work by focusing on the dark side of the intersection between immersive technology and cultural heritage, without addressing the ethical issues underlying the development of either technology.

There is a clear need for urgent strategies aimed not only at harnessing the positive potential of these technologies, but also at mitigating their adverse effects. This study will focus on identifying and developing such strategies, with an emphasis on improving the efficacy of immersive technologies in cultural heritage, while addressing the ethical, social and technological challenges they pose. This approach will contribute to a more nuanced understanding of the role of digital technologies in cultural preservation and engagement, providing a balanced perspective that can guide future implementation and policy development.

\section{Method}

With the aim of identifying and examining the latest status of research in the digital applications of cultural heritage up to 2024, we chose to conduct a scoping review. The purpose of the scope review is to ``\textit{provide an initial delineation and assessment of the potential size and coverage of the existing research literature}''. Although the issues concerned in systematic reviews are usually relatively precise, for example, the existing reviews in relevant fields usually focus on the current status of the application of cultural heritage in a specific field, which is a scope-defining review, lacking a comprehensive and systematic research perspective exploration that includes all the applications of cultural heritage technologies. They will clearly present and organize the relevant field evidence found, and analyze existing research and contributions under specific topics. In this paper, we conduct research that is up-to-date in time and relatively larger in scope, rather than proposing solutions to specific problems. For this reason, we limit our literature screening to the full text of papers and scope review of the technologies, applications and devices of cultural heritage.

\subsection{Research Questions and Rationale}

This review is guided by the following research questions (RQs): 
\begin{itemize} 
\item What types of equipment are primarily used, and what technologies are commonly employed when interacting with digital displays of cultural heritage? 
\begin{itemize} 
\item For example, are VR headsets or AR glasses more commonly used in museum installations? 
\item What role do motion sensors or interactive touchscreens play in enhancing user engagement with cultural heritage artifacts? 
\end{itemize} 
\item What are the key application areas developed in this context? 
\begin{itemize} 
\item For instance, are these technologies primarily applied in education, tourism, or preservation of intangible heritage such as traditional crafts or oral history? 
\end{itemize} 
\item What potential challenges or negative impacts arise from the use of immersive technologies in cultural heritage? 
\begin{itemize} 
\item Examples could include ethical concerns over the representation of sensitive cultural materials or the long-term preservation of digital installations themselves. 
\item Additionally, there may be technical limitations such as high costs or a lack of accessibility for certain audiences. 
\end{itemize} 
\end{itemize}

\subsection{Search Strategy}

\subsubsection{Identify Keywords}
At the beginning of the search, we tested the search results by browsing a large number of relevant articles on Google Scholar and trying to use some high-frequency words as keywords. Notably, to avoid subjective interference, we chose to be data-driven by searching for roots such as ``\textsc{immersion}'' and ``\textsc{interaction}.'' Therefore, ``\textsc{Cultural Heritage}'' AND ``\textsc{Techn*}'' AND (``\textsc{Digital}'' OR \textsc{Immersi*} OR ``\textsc{Multimedia}'' OR \textsc{Interacti}*) were finally selected as keywords for this review. This means that we will have access to most of the data related to this set of words without bias in retrieval due to our linguistic predisposition and knowledge background.

After defining the keywords related to the digitization of cultural heritage, we searched through the keywords. ACM Digital Library, IEEE Xplore and Scopus databases are widely used in data retrieval of core papers and have gained high evaluation in the industry. Therefore, we used the above databases to retrieve 339, 520 and 4,509 articles on ACM Digital Library, IEEE Xplore and Scopus, accordingly (5,368 articles in total).

\subsubsection{Search}
We searched ACM Digital Library, IEEE Xplore, Scopus databases using the OR operator and the AND operator between keywords within each set. Since the search form of each database is different, we give the same keyword format of the three databases slightly different when searching. We limit our search to an article's title, abstract, and author keywords.

\subsubsection{Preliminary Literature Format Screening}
Before the first round of screening, we first removed some records that did not meet the criteria from the retrieved papers, including records that were flagged as unqualified by the automated tool, duplicate records, and records that were deleted for other reasons. The purpose of this process is to improve the efficiency of data refinement screening by screening and removing targets that do not meet the criteria. Then, we use the filter to search the database and get ACM Digital Library (205), IEEE Xplore (513), and Scopus Search (3,822). The search logic we use is to remove papers of low value or interference such as posters, short papers, abstract papers, and copyright notices.

\subsection{First Screening}
After combining the search results from the three databases, we obtained 3,907 articles as a sample for the first round of screening. For these objects, we use Excel tables for manual screening, which is divided into three stages of the screening process. In the first stage, we screened the articles for language and excluded 1,836 articles written in non-English. Of the 2,071 surviving papers, we have only retained 1,448 journal papers. After checking 623 English conference papers, we found problems such as changing the name of the conferences, co-organizing the conferences, and stopping the conferences. For this reason, we believe that research based on conference papers is not very reliable, so journal papers are listed as the main body of our research. In the third stage, 382 journal papers were retained by screening papers based on page count, impact factor, citations, word of mouth, retractions, or duplication factors.

\subsection{Filtering by Article Content}
In this stage, we mainly screened the full-text content. We condensed the 382 papers to 177 according to three evaluation criteria: term application, keyword correlation, and content coverage, and entered the final systematic study. Figure~\ref{fig:Filtering by article content} details the complete process from launching the search to identifying the papers included in the analysis. Based on our research questions, we defined the following exclusion (EC) and inclusion (IC) criteria:

\begin{figure}[ht]
    \centering
    \includegraphics[width=0.95\linewidth]{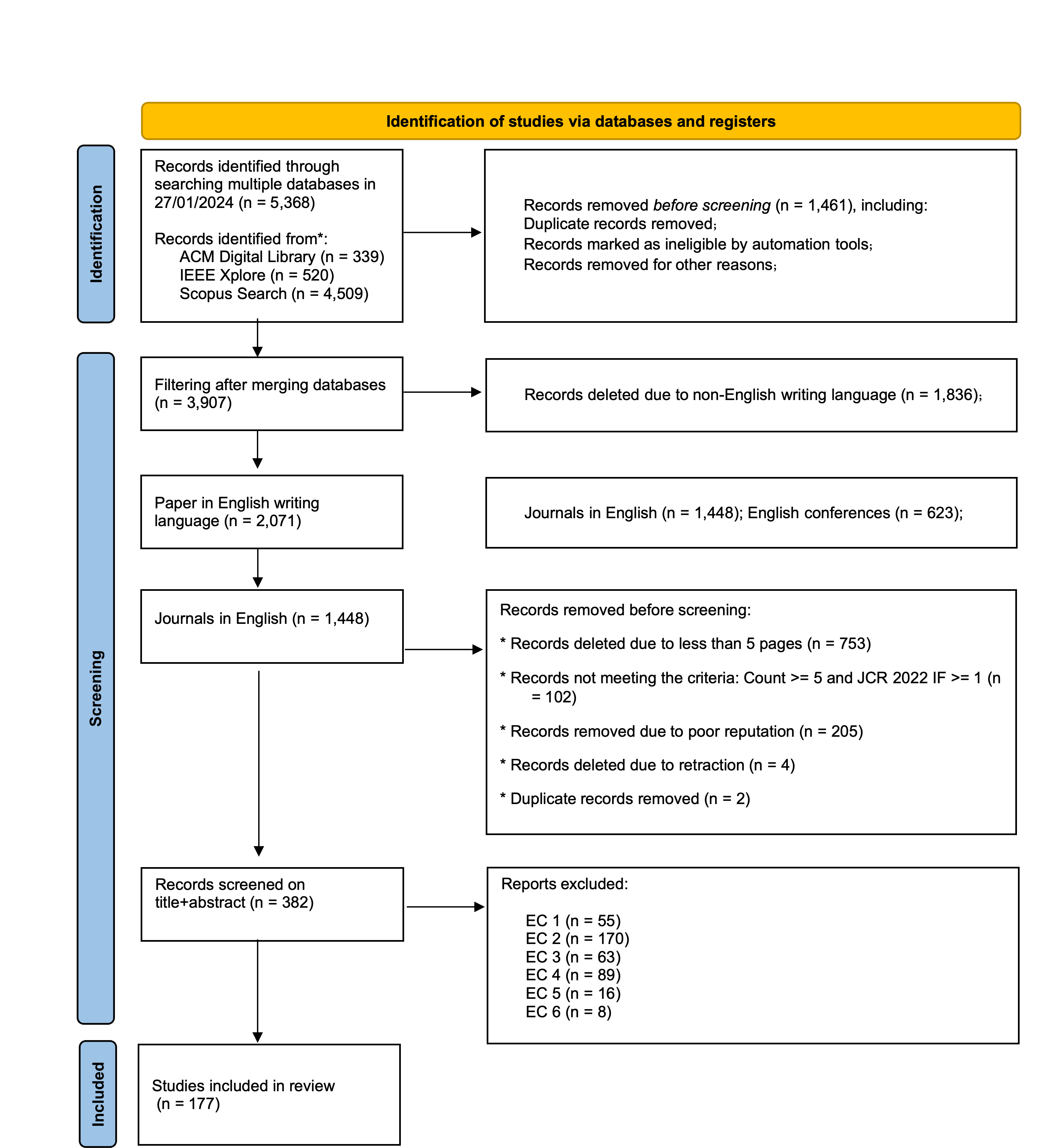}
    \caption{PRISMA flow diagram illustrating the identification, screening, and inclusion process of studies for a systematic review. Initially, 5,368 records were identified through searches in multiple databases (ACM Digital Library, IEEE Xplore, and Scopus). After the removal of 1,461 duplicates and non-relevant entries, 3,907 records were further filtered. Records written in non-English languages were excluded (n = 1,836), leaving 2,071 papers. A total of 1,448 journal papers and conference proceedings were screened, and additional exclusions were made based on criteria such as page length, impact factor, and paper quality. 177 studies were finally included in the review.}
    \label{fig:Filtering by article content}
\end{figure}

\textbf{EC1.} \textit{Reviews or literature reviews}: Reviews, literature reviews, and opinion articles are excluded. The purpose of this study is to analyze specific cases and original research, not to summarize or review existing review studies.

\textbf{EC2.} \textit{Key terms missing}: If relevant key terms (e.g., AR, VR, immersive technologies, etc.) are not mentioned in the article, they are excluded. This reflects that the article may not be relevant to the research topic.

\textbf{EC3.} \textit{Inappropriate Use of Terminology}: If the description or application of digital technology in the article is not consistent with the context in which cultural heritage is displayed, it is excluded. For example, if the ``metadata'' discussed in the article is used only for general information management or data storage, and does not relate to its application in the presentation of cultural heritage, such as to enhance the accessibility of cultural heritage materials, improve the interactivity of content or enrich the visitor experience, then such use is considered an inappropriate use of the term. Because it does not properly reflect the important role and potential value of metadata in the display of cultural heritage. Similarly, if the technology is used for gamified learning or the simple presentation of cultural heritage content without an in-depth discussion of its specific application and impact in the presentation of cultural heritage, the article does not meet the research criteria.

\textbf{EC4.} \textit{Lack of correlation between digital technology and cultural heritage display}: If the application of digital technology in the field of cultural heritage display is not explicitly discussed in the article, even if digital technology (e.g., AR, VR, etc.) is mentioned, it will be excluded. This is because a reference to digital technology without exploring its specific application and impact in the display of cultural heritage does not meet the core concerns of this study. For example, articles that focus on the protection of cultural heritage rather than its display, or discuss technologies that have a direct link to the display of cultural heritage, will also be excluded.

\textbf{EC5.} \textit{Marginal references to digital technology}: Articles will be excluded if the reference to digital technology is only superficial, such as a brief mention in the abstract or introduction, or if the application of digital technology is weakly linked to the display of cultural heritage and does not constitute the main content of the article. This criterion aims to ensure that the included articles focus on the practical application and in-depth analysis of digital technologies in the presentation of cultural heritage.

\textbf{EC6.} \textit{Data set introduction}: If the article mainly introduces data sets but does not explicitly mention immersive or emerging technologies, the article will be excluded.

\textbf{EC7.} \textit{Lack of information}: An article will be excluded if it does not provide enough detailed information to allow the full application of the coding standard.

\textbf{IC1.} \textit{Innovative applications of technology}: Articles should show how emerging technologies (e.g., artificial intelligence, AR, VR, etc.) can be applied to the presentation and experience of cultural heritage. This includes, but is not limited to, innovative applications of technology in cultural heritage, ways to enhance interactive experiences, and the application of new technologies in cultural heritage education.

\textbf{IC2.} \textit{Integration of technology and cultural heritage}: Research should explore new ways of integrating technology and cultural heritage, such as recreating historical scenes through AR, or using VR to display cultural sites that are not physically accessible.

\textbf{IC3.} \textit{User Interaction and Experience}: includes studies that focus on how users interact with cultural heritage enhanced by technology. This may involve the evaluation of user experience, innovation in interactive design, or research into how users perceive and understand cultural heritage through technology.

\textbf{IC4.} \textit{Application-oriented research}: Consider those studies in which the application of technology is not a primary research focus but plays a key role in the presentation of cultural heritage. For example, a project that uses AR to aid history education, or a study that applies machine learning techniques to analyze cultural heritage data.

\textbf{IC5.} \textit{Articles requiring further reading}: If, at the time of initial reading of the abstract, it is not clear whether the articles meet the inclusion criteria, we will include them for the second stage of screening.

For the process of the literature screening, we employed a multi-step approach that included independent screening by multiple authors, as well as a review by a panel of 20 experts. These experts, comprising cultural heritage presentation practitioners, university scholars, and professionals from cultural heritage protection institutions, were invited to assess and discuss the initial selection results. Their involvement enhances both the objectivity and credibility of the screening process. Additionally, an anonymized table (see \autoref{tab:commands2}) will be included in the manuscript, providing background information on these 20 experts, such as their role, gender, years of experience (categorized as senior [$>15$ years] or junior [$<5$ years]), and professional background (i.e., academic, industry, and public sector). This information is included to ensure greater transparency regarding the composition and expertise of the panel.

\begin{sidewaystable}[htbp]
    \centering 
    \caption{Anonymized Background Information of the 20 Experts in the Review Panel}
    \label{tab:commands2}
     \scriptsize 
     \begin{tabular}{p{2cm}p{1cm}p{2cm}p{7cm}p{2cm}p{2cm}} 
    \toprule
    Expert & Gender & Age & Role & Background & Experience\\
    \midrule

\verb|1| & Male & 55 & Professor & Academic & Senior\\
\verb|2| & Female & 36 & PhD Student & Academic & Senior\\
\verb|3| & Female & 31 & Historian & Academic & Junior\\
\verb|4| & Male & 33 & Architectural Restorer & Academic & Junior\\
\verb|5| & Male & 46 & Cultural Creative Company Owner & Industry & Senior\\
\verb|6| & Female & 41 & Heritage Site Manager & Public Sector & Senior\\
\verb|7| & Male & 34 & Heritage Site Interpreter & Public Sector & Senior\\
\verb|8| & Male & 29 & Local Museum Staff & Public Sector & Junior\\
\verb|9| & Male & 30 & National Museum Staff & Public Sector & Junior\\
\verb|10| & Female & 35 & Cultural Heritage Column Website Manager & Industry & Junior\\
\verb|11| & Female & 36 & Cultural Heritage Column Reporter & Industry & Senior\\
\verb|12| & Male & 37 & Associate Professor & Academic & Senior\\
\verb|13| & Male & 38 & Local Museum Staff & Public Sector & Senior\\
\verb|14| & Male & 44 & Vice President of Cultural Creative Group & Industry & Senior\\
\verb|15| & Male & 45 & Local Museum Staff & Public Sector & Senior\\
\verb|16| & Female & 40 & Local Museum Staff & Public Sector & Senior\\
\verb|17| & Male & 44 & Local Museum Staff & Public Sector & Senior\\
\verb|18| & Female & 35 & Local Museum Staff & Public Sector & Junior\\
\verb|19| & Male & 40 & Researcher & Academic & Senior\\
\verb|20| & Female & 33 & Local Museum Staff & Public Sector & Junior\\

    \bottomrule
    \end{tabular}
\end{sidewaystable}

\subsection{Critical Appraisal, Potential Bias, and Limitations}

The scope of the sample should be limited by a more comprehensive consideration of all available types of evidence in content, without limiting the breadth of evidence sources due to methodological quality. However, in order to obtain a relatively more manageable set of papers, we included only complete thesis proceedings and journal publications in the final sample range. Apart from these screenings, we did not exclude papers based on their methodological quality.

The keyword determination process may be influenced by the specific papers in the research base. In addition, it is difficult for the keywords we have chosen to cover all articles in the relevant field, so the existing list may carry a subjective understanding of this field of study. But because we use root search, which avoids the limitations of subjectivity to some extent, we are confident that we can find most relevant papers. But this approach also leaves us with a large sample of false positives. To reduce errors in the screening and coding stages, we used the first round of a three-stage screening process to eliminate as many errors as possible. Our aim is to cover the latest extensive research in the field of digital display of cultural heritage. In addition, we conducted extensive discussion sessions to resolve conflicts and fine-track exclusion/inclusion criteria, keyword descriptions, screening processes (two one-hour initial screening sessions, three one-hour full-text screening and data extraction sessions).

\section{Results}

\begin{sidewaysfigure}[htbp]
    \centering
    \includegraphics[width=\linewidth]{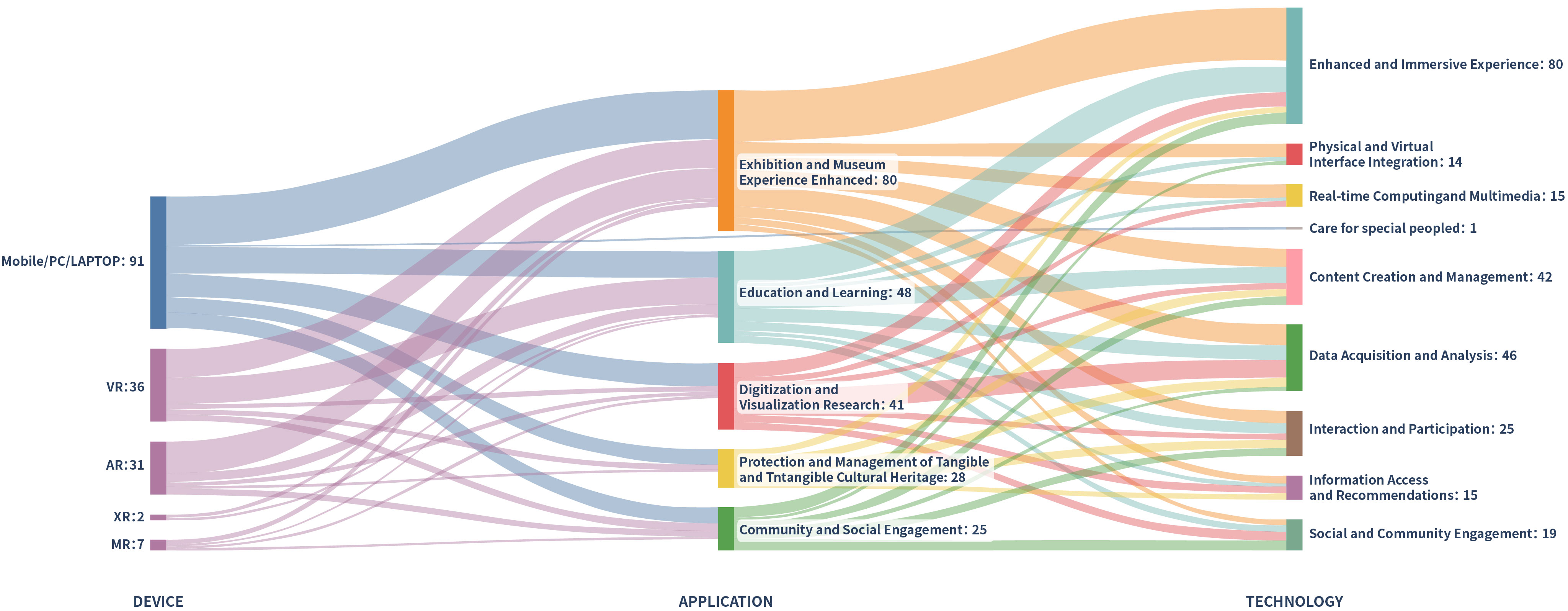}
    \caption{Sankey Diagram of Device, Application and Technology}
    \label{fig:sankey}
\end{sidewaysfigure}

\subsection{Paper Review}

\subsubsection{Research Direction}
Based on the final corpus of 177 unique papers, we found four research directions. First of all, the largest proportion of papers are those with a specific research project as the core research object. These papers focus on the application of cultural heritage digitization strategies in specific scenarios and aim to explore a comprehensive analytical framework for practical application. For example, by building a digital platform for cultural heritage, HeGO, users can use their uploaded images to build three-dimensional reconstructions of sites and apply them to implement social media games~\citep{fontanella_hego_2021}.

Second, methodological framework research, mainly discusses the solution of a kind of cultural heritage digital display problems. For example, Suominen et al propose an approach to play culture's internal and external cultural heritage, discussing the relationship between history, cultural heritage and digital technology through a four-fold table of the relationship between cultural heritage (or history) and digital technology~\citep{suominen_gaming_2013}. Third, discuss the application of a new technology in the application of multiple projects. Ardito et al., for example, propose ways to identify and specify the attributes of Internet of Things (IoT) devices through user-defined semantics in order to make it easier for cross-domain experts to operate the technology and apply IoT technology more smoothly in the field of cultural heritage~\citep{ardito_user-defined_2020}. Fourth, the smallest proportion of literature review studies. Hou et al., for example, outline the models, techniques, and practices that drive the digitized life of intangible cultural heritage resources, looking at several key but less researched tasks such as digital archiving, computational coding, conceptual representation, and interactive engagement with intangible cultural elements to identify advancements and gaps in existing conventions~\citep{hou_digitizing_2022}.

\subsubsection{Publication Venue}

A focus on publication venues can quickly determine the distribution of core journals within the field. According to the cumulative number of published papers, these papers were published in several publishing places, among which the most common five were \textit{Journal on Computing and Cultural Heritage} (35), \textit{Journal of Cultural Heritage} (22), \textit{Digital Applications in Archaeology and Cultural Heritage} (12), \textit{Personal and Ubiquitous Computing} (12), and \textit{Heritage Science} (11). 

\subsubsection{Keyword Distribution}

\subsection{Visual Analysis}

\subsubsection{Sample Classification Analysis}
We have carefully read and studied the full text of the 177 papers selected finally. We studied these papers according to three main categories: ``device'', ``application'', and ``technology''.

\textbf{Device} refers to the devices used by the audience in the usage scenario mentioned in the paper. For example, Hulusic mentioned in the article that Tangible User Interfaces (TUI), in which users interact with the virtual cultural heritage application through the actual operation of the interface, can be summarized in the ``mobile/PC/LAPTOP'' category~\citep{hulusic_tangible_2023}. According to the data analysis, we mainly extracted a total of 199 frequency articles mentioned about equipment in the article. These include AR and VR devices, touchable and interactive user interface devices, mobile and portable devices, data acquisition scanning and printing devices, sensors and computing devices, projection and mapping technology devices, photography, and video capture. Among them, the most widely used are AR and VR devices, which are mentioned 58 times, accounting for 29.1\%. VR devices account for the largest proportion. In addition, data acquisition, 3D scanning and printing equipment were also mentioned 45 times, accounting for 22.6 percent. 3D laser scanners and photogrammetry equipment were used 22 times. In contrast, there are fewer articles that mention the use of touchable and interactive user interfaces, projection and mapping technology devices, photography and video capture devices, and we believe that there is still a lot of room for development in this field of devices (see \autoref{tab:commands3}).

\begin{sidewaystable}[htbp]
    \centering 
    \caption{Categorization of Devices Mentioned in Articles on Immersive Cultural Heritage Interaction Experiences}
    \label{tab:commands3}
     \scriptsize 
     \begin{tabular}{p{3cm}p{14cm}} 
    \toprule
    Device & Paper\\
    \midrule

    \verb|AR| &~\citep{echavarria_creative_2022, chiang_augmented_2023, ridel_revealing_2014, damala_musetech_2019, jang_content_2023, odwyer_volumetric_2021, hu_individually_2023, marcos_digital_2007, sertalp_raising_2023, zhu2024designing, arayaphan2022digitalization, gatelier_business_2022, stichelbaut_towards_2021, malik_3d_2021, lassandro_analysing_2021, petrovic_geodetic_2021, zhang_multimedia-based_2021, cejka_hybrid_2020, bozzelli_integrated_2019, andrade_phygital_2020, chng_crowdsourcing_2019, Massimo_perceiving_nodate, villa_lapidary_2017, szabo_landscape_2017, brancati_experiencing_2017, younes_pose_2016, comes_methodology_2014, chen_case_2013, carrozzino_virtually_2011, neamtu_evaluating_2024, krumpen_towards_2021} \\[6pt]
    
    \verb|MR| &~\citep{millard_balance_2020, hammady_ambient_2020, trunfio_experimenting_2023, zaia2022egyptian, bekele_walkable_2019, rahaman_photo_2019, jacobs_classification_2006} \\[6pt]
    
    \verb|VR| &~\citep{argyriou_design_2020, han_compelling_2020, han_compelling_2020, han_compelling_2020, wenzhi_chen_animations_2010, hu_individually_2023, petrelli_exploring_2023, tong_digital_2023, lu_development_2023, zhao2023personalized, bajaj2024design, wu_design_2022, arayaphan2022digitalization, gatelier_business_2022, mas_digitization_2022, monaco_linked_2022, leow_analysing_2021, malik_3d_2021, lassandro_analysing_2021, ocon_digitalising_2021, ghani_effect_2020, zhang_multimedia-based_2021, ferdani_3d_2020, loaiza_carvajal_virtual_2020, bozzelli_integrated_2019, chng_crowdsourcing_2019, kersten_historic_2018, serain_sensitive_2018, Massimo_perceiving_nodate, brumana_virtual_2018, chen_case_2013, carrozzino_virtually_2011, bruno_3d_2010, lutz_virtual_1999, neamtu_evaluating_2024, krumpen_towards_2021} \\[6pt]
    
    \verb|XR| &~\citep{zaia2022egyptian, neamtu_evaluating_2024} \\[6pt]
    
    \verb|Mobile/PC/Laptop| &~\citep{tsiviltidou_digital_2023, hulusic_tangible_2023, viola_networks_2023, semeraro_folksonomy-based_2012, not_digital_2019, froschauer_art_2013, green_use_2021, fontanella_hego_2021, diaz_imanote_2011, pietroni_interacting_2014, damala_musetech_2019, konstantakis_adding_2020, suominen_gaming_2013, zhao_supporting_2020, jones_enthusiast_2019, hess_developing_2015, montusiewicz_architectural_2021, rattanarungrot_preserving_2023, drygalska_technologically_2023, petrelli_exploring_2023, li_kano-qfd-based_2023, qi_application_2024, renzi_storytelling_2023, xiong_titul_2023, pistofidis2023design, zhu2024designing, ma_spreading_2023, wu_design_2022, aburamadan_developing_2022, chen_development_2022, arayaphan2022digitalization, zhenrao_joint_2021, kuna_interactive_2022, portales_virtual_2021, simone_museums_2021, spagnolo_fringe-projection_2021, de_lazaro_shedding_2021, petrovic_geodetic_2021, ardito_user-defined_2020, cejka_hybrid_2020, permatasari_web_2020, quattrini_digital_2020, bhaumik_conserving_2020, luznik-jancsary_integration_2020, jaillot_describing_2020, zhou_database_2020, marsili_digital_2019, andrade_phygital_2020, tai_digital_2020, machidon_culturalerica_2020, ricart_promoting_2019, deng_selective_2019, bruno_enhancing_2019, jiang_cloth_2019, vurpillot_aspectus_2019, gil-meliton_historical_2019, dalbertis_encyclopedic_2018, smith_assessing_2018, noardo_architectural_2018, gonzalez_zarandona_digitally_2018, Massimo_perceiving_nodate, solima_qr_2018, kuroczynski_virtual_2017, rojas-sola_digital_2018, villa_lapidary_2017, kim_ontology-based_2017, smirnov_context-based_2017, bujari_using_2017, scopigno_delivering_2017, celentano_layered_2017, carletti_participatory_2016, civantos_using_2016, aigner_heritage-making_2016, bartolini_recommending_2016, vosinakis_visitor_2016, peters_digital_2015, li_interactive_2015, liritzis_digital_2015, liew_towards_2014, grainger_clemson_trailfinderscurating_2014, dimoulas_audiovisual_2014, smith_participatory_2014, carrozzino_designing_2013, reerink_unicum_2012, carrozzino_virtually_2011, giaccardi_social_2008, styliadis_metadata-based_2009, gutierrez_ai_2007, arlitsch_utah_2003, neudecker_user_2010, leonov_laser_2015} \\[6pt]
    
    \verb|None| &~\citep{psomadaki_digital_2019, li_visitors_2023, smithies_madih_2023, hou_digitizing_2022, katifori_let_2020, garcia-molina_digitalization_2021, raheb_moving_2021, el-hakim_detailed_2004, vilbrandt_digitally_2011, balzani_saving_2024, li_scene_2023, xu_construction_2023, yan_intelligent_2023, zhu_designing_2024, fan_research_2023, wu_digital_2023, staropoli_reflections_2023, ocon2023low, cheng_using_2022, li_comparative_2021, deligiorgi_3d_2021, kamariotou_strategic_2021, croix_recasting_2020, wijnhoven_digital_2020, nancarrow_countering_2019, xue_evaluation_2019, yaagoubi_seh-sdb_2019, mah_generating_2019, liang_semantic-based_2019, adembri_virtual_2018, hong_social_2017, cheok_confucius_2017, themistocleous_model_2017, sutcliffe_understanding_2014, georgopoulos_3d_2014, menna_hybrid_2014, yastikli_documentation_2007} \\[6pt]

    \bottomrule
    \end{tabular}
\end{sidewaystable}

\textbf{Application} refers to the application field after the project development mentioned in the paper. For example, Tsiviltidou's article explores the case of using digital storytelling in the classroom to construct inquiry-based learning with digital museum collections, a project applied to classroom instruction and thus grouped under the type of application ``Education and learning''~\citep{tsiviltidou_digital_2023}. According to the data analysis, we mainly extracted a total of 195 applications mentioned in the article, including education and learning, community and social participation, digitization and visualization research, virtual and AR technology applications, exhibition and museum experience enhancement, material and intangible cultural heritage protection and management, care for special groups, a total of 7 application fields. It is worth noting that 61 articles mentioned the field of exhibition and museum experience enhancement, accounting for 31.2\%. There is a similar performance in the field of education and learning, accounting for 24.1 percent. It can be found that the application direction of the digital display of cultural heritage is still mainly to improve the basic function of user experience, while helping users better understand and learn the exhibition content, which is also the responsibility of museums. In addition, it is worth considering that there are relatively few applications in the field of community and social engagement, accounting for 9\%. Therefore, we believe that in the process of digital display of cultural heritage, we should further open up the integration with the society and give play to the social functions of museums. In addition, the protection of cultural heritage and the care of special groups are the areas with the least attention, which also reflects the gap in the application of cultural heritage digitization in this field to a certain extent (see \autoref{tab:commands4}).

\begin{sidewaystable}[htbp]
    \centering 
    \caption{Categorization of Applications Mentioned in Articles on Immersive Cultural Heritage Interaction Experiences}
    \label{tab:commands4}
    \scriptsize 
    \begin{tabular}{p{11cm}p{6cm}} 
    \toprule
    Application & Paper\\
    \midrule

    \verb|Education and Learning| &~\citep{echavarria_creative_2022, tsiviltidou_digital_2023, chiang_augmented_2023, hulusic_tangible_2023, green_use_2021, argyriou_design_2020, psomadaki_digital_2019, pietroni_interacting_2014, wenzhi_chen_animations_2010, rattanarungrot_preserving_2023, el-hakim_detailed_2004, lu_development_2023, zhao2023personalized, wu_design_2022, arayaphan2022digitalization, mas_digitization_2022, li_comparative_2021, leow_analysing_2021, spagnolo_fringe-projection_2021, de_lazaro_shedding_2021, ghani_effect_2020, ferdani_3d_2020, loaiza_carvajal_virtual_2020, bhaumik_conserving_2020, bhaumik_conserving_2020, marsili_digital_2019, deng_selective_2019, bruno_enhancing_2019, jiang_cloth_2019, kersten_historic_2018, serain_sensitive_2018, villa_lapidary_2017, kim_ontology-based_2017, brumana_virtual_2018, szabo_landscape_2017, cheok_confucius_2017, carletti_participatory_2016, civantos_using_2016, vosinakis_visitor_2016, sutcliffe_understanding_2014, menna_hybrid_2014, chen_case_2013, carrozzino_designing_2013, reerink_unicum_2012, styliadis_metadata-based_2009, arlitsch_utah_2003, neamtu_evaluating_2024, krumpen_towards_2021} \\[6pt]

    \verb|Community and Social Engagement| &~\citep{echavarria_creative_2022, chiang_augmented_2023, froschauer_art_2013, fontanella_hego_2021, psomadaki_digital_2019, smithies_madih_2023, millard_balance_2020, katifori_let_2020, jones_enthusiast_2019, montusiewicz_architectural_2021, wenzhi_chen_animations_2010, rattanarungrot_preserving_2023, marcos_digital_2007, ferdani_3d_2020, jaillot_describing_2020, machidon_culturalerica_2020, chng_crowdsourcing_2019, dalbertis_encyclopedic_2018, noardo_architectural_2018, gonzalez_zarandona_digitally_2018, carletti_participatory_2016, aigner_heritage-making_2016, grainger_clemson_trailfinderscurating_2014, chen_case_2013, neudecker_user_2010} \\[6pt]

    \verb|Digitization and Visualization Research| &~\citep{viola_networks_2023, fontanella_hego_2021, diaz_imanote_2011, drygalska_technologically_2023, el-hakim_detailed_2004, balzani_saving_2024, li_scene_2023, xu_construction_2023, qi_application_2024, xiong_titul_2023, zhu_designing_2024, fan_research_2023, ocon2023low, chen_development_2022, cheng_using_2022, li_comparative_2021, deligiorgi_3d_2021, spagnolo_fringe-projection_2021, zhang_multimedia-based_2021, croix_recasting_2020, quattrini_digital_2020, tai_digital_2020, ricart_promoting_2019, rahaman_photo_2019, liang_semantic-based_2019, chng_crowdsourcing_2019, vurpillot_aspectus_2019, gil-meliton_historical_2019, adembri_virtual_2018, serain_sensitive_2018, smith_assessing_2018, kuroczynski_virtual_2017, rojas-sola_digital_2018, smirnov_context-based_2017, bujari_using_2017, scopigno_delivering_2017, georgopoulos_3d_2014, carrozzino_virtually_2011, giaccardi_social_2008, jacobs_classification_2006, yastikli_documentation_2007} \\[6pt]

    \verb|Exhibition and Museum Experience Enhanced| &~\citep{ridel_revealing_2014, not_digital_2019, carrozzino_beyond_2010, li_visitors_2023, damala_musetech_2019, millard_balance_2020, hammady_ambient_2020, jang_content_2023, odwyer_volumetric_2021, hess_developing_2015, hu_individually_2023, drygalska_technologically_2023, petrelli_exploring_2023, tong_digital_2023, renzi_storytelling_2023, sertalp_raising_2023, zhao2023personalized, zhu2024designing, wu_digital_2023, bajaj2024design, trunfio_experimenting_2023, arayaphan2022digitalization, zaia2022egyptian, gatelier_business_2022, monaco_linked_2022, zhenrao_joint_2021, stichelbaut_towards_2021, portales_virtual_2021, kamariotou_strategic_2021, simone_museums_2021, de_lazaro_shedding_2021, lassandro_analysing_2021, ocon_digitalising_2021, petrovic_geodetic_2021, ghani_effect_2020, ardito_user-defined_2020, zhang_multimedia-based_2021, cejka_hybrid_2020, quattrini_digital_2020, zhou_database_2020, bozzelli_integrated_2019, andrade_phygital_2020, bekele_walkable_2019, yaagoubi_seh-sdb_2019, mah_generating_2019, liang_semantic-based_2019, bruno_enhancing_2019, jiang_cloth_2019, gil-meliton_historical_2019, adembri_virtual_2018, dalbertis_encyclopedic_2018, serain_sensitive_2018, noardo_architectural_2018, Massimo_perceiving_nodate, solima_qr_2018, rojas-sola_digital_2018, villa_lapidary_2017, brancati_experiencing_2017, scopigno_delivering_2017, celentano_layered_2017, carletti_participatory_2016, civantos_using_2016, aigner_heritage-making_2016, younes_pose_2016, bartolini_recommending_2016, peters_digital_2015, li_interactive_2015, liritzis_digital_2015, liew_towards_2014, comes_methodology_2014, georgopoulos_3d_2014, smith_participatory_2014, bruno_3d_2010, giaccardi_social_2008, styliadis_metadata-based_2009, gutierrez_ai_2007, arlitsch_utah_2003, lutz_virtual_1999, neamtu_evaluating_2024, leonov_laser_2015} \\[6pt]

    \verb|Management of (In)Tangible Cultural Heritage| &~\citep{semeraro_folksonomy-based_2012, green_use_2021, diaz_imanote_2011, han_compelling_2020, hou_digitizing_2022, zhao_supporting_2020, hess_developing_2015, garcia-molina_digitalization_2021, raheb_moving_2021, vilbrandt_digitally_2011, balzani_saving_2024, xu_construction_2023, yan_intelligent_2023, li_kano-qfd-based_2023, xiong_titul_2023, zhu_designing_2024, fan_research_2023, staropoli_reflections_2023, ma_spreading_2023, aburamadan_developing_2022, arayaphan2022digitalization, kuna_interactive_2022, malik_3d_2021, permatasari_web_2020, nancarrow_countering_2019, xue_evaluation_2019, themistocleous_model_2017, dimoulas_audiovisual_2014} \\[6pt]

    \verb|Care for special peopled| &~\citep{pistofidis2023design} \\[6pt]

    \raggedright \verb|None| &~\citep{konstantakis_adding_2020, suominen_gaming_2013, wijnhoven_digital_2020, hong_social_2017} \\[6pt]

    \bottomrule
    \end{tabular}
\end{sidewaystable}

\textbf{Technology} refers to the technology used in the project development process mentioned in the paper. For example, Echavarria et al. described the presentation of different cultural heritage stories through interactive AR maps that use AR technology to enhance physical elements, which can be grouped under the category ``Technologies for enhanced and immersive experiences''~\citep{echavarria_creative_2022}. According to the data analysis, we mainly extracted a total of 311 frequency of articles mentioning equipment in the article. It includes eight technology types: enhancement and immersion experience technology, interactive participation technology, social and community participation technology, content creation and management technology, data acquisition and analysis technology, information access and recommendation technology, physical and virtual interface integration technology, real-time computing and multimedia technology. Among them, the highest frequency of use is enhanced and immersive experience technology, up to 96 times, accounting for 30.9\%. This kind of technology can be more inclined to meet the user's sense of experience and create more emotional force in the same time and space, so it has been invested in more attention. In addition, social and community engagement, content creation and management, and data acquisition and analysis are three areas where considerable attention has been devoted (see \autoref{tab:commands5}).

\begin{sidewaystable}[htbp]
    \centering 
    \caption{Categorization of Technology Mentioned in Articles on Immersive Cultural Heritage Interaction Experiences}
    \label{tab:commands5}
    \scriptsize 
    \begin{tabular}{p{11cm}p{6cm}} 
    \toprule
    Application & Paper\\
    \midrule

    \verb|Enhanced and Immersive Experience| &~\citep{echavarria_creative_2022, chiang_augmented_2023, hulusic_tangible_2023, ridel_revealing_2014, argyriou_design_2020, han_compelling_2020, carrozzino_beyond_2010, li_visitors_2023, damala_musetech_2019, millard_balance_2020, hammady_ambient_2020, hou_digitizing_2022, jang_content_2023, odwyer_volumetric_2021, wenzhi_chen_animations_2010, hu_individually_2023, petrelli_exploring_2023, marcos_digital_2007, el-hakim_detailed_2004, tong_digital_2023, lu_development_2023, xiong_titul_2023, zhao2023personalized,zhu2024designing, bajaj2024design, staropoli_reflections_2023, trunfio_experimenting_2023, wu_design_2022, arayaphan2022digitalization, zaia2022egyptian, gatelier_business_2022, mas_digitization_2022, li_comparative_2021, monaco_linked_2022, stichelbaut_towards_2021, deligiorgi_3d_2021, kamariotou_strategic_2021, simone_museums_2021, leow_analysing_2021, malik_3d_2021, lassandro_analysing_2021, ocon_digitalising_2021, petrovic_geodetic_2021, ghani_effect_2020, zhang_multimedia-based_2021, cejka_hybrid_2020, ferdani_3d_2020, loaiza_carvajal_virtual_2020, bozzelli_integrated_2019, andrade_phygital_2020, bekele_walkable_2019, rahaman_photo_2019, bruno_enhancing_2019, jiang_cloth_2019, chng_crowdsourcing_2019, kersten_historic_2018, serain_sensitive_2018, Massimo_perceiving_nodate, kuroczynski_virtual_2017, villa_lapidary_2017, kim_ontology-based_2017, szabo_landscape_2017, scopigno_delivering_2017, aigner_heritage-making_2016, younes_pose_2016, grainger_clemson_trailfinderscurating_2014, comes_methodology_2014, menna_hybrid_2014, chen_case_2013, carrozzino_designing_2013, bruno_3d_2010, giaccardi_social_2008, styliadis_metadata-based_2009, gutierrez_ai_2007, jacobs_classification_2006, arlitsch_utah_2003, lutz_virtual_1999, neamtu_evaluating_2024, krumpen_towards_2021, leonov_laser_2015}
 \\[6pt]

    \verb|Interaction and Participation| &~\citep{viola_networks_2023, not_digital_2019, froschauer_art_2013, green_use_2021, han_compelling_2020, psomadaki_digital_2019, damala_musetech_2019, hou_digitizing_2022, zhao_supporting_2020, hess_developing_2015, raheb_moving_2021, drygalska_technologically_2023, bajaj2024design, ma_spreading_2023, wu_design_2022, leow_analysing_2021, ferdani_3d_2020, bozzelli_integrated_2019, mah_generating_2019, cheok_confucius_2017, li_interactive_2015, sutcliffe_understanding_2014, smith_participatory_2014, carrozzino_virtually_2011, leonov_laser_2015} \\[6pt]

    \verb|Social and Community Engagement| &~\citep{semeraro_folksonomy-based_2012, froschauer_art_2013, fontanella_hego_2021, psomadaki_digital_2019, damala_musetech_2019, suominen_gaming_2013, wenzhi_chen_animations_2010, rattanarungrot_preserving_2023, wu_design_2022, bozzelli_integrated_2019, rahaman_photo_2019, vurpillot_aspectus_2019, kuroczynski_virtual_2017, bujari_using_2017, carletti_participatory_2016, smith_participatory_2014, chen_case_2013, neudecker_user_2010, leonov_laser_2015} \\[6pt]

    \verb|Content Creation and Management| &~\citep{tsiviltidou_digital_2023, hulusic_tangible_2023, diaz_imanote_2011, psomadaki_digital_2019, smithies_madih_2023, pietroni_interacting_2014, damala_musetech_2019, hou_digitizing_2022, katifori_let_2020, marcos_digital_2007, li_scene_2023, yan_intelligent_2023, qi_application_2024, sertalp_raising_2023, xiong_titul_2023, ocon2023low, ma_spreading_2023, chen_development_2022, arayaphan2022digitalization, kuna_interactive_2022, stichelbaut_towards_2021, kamariotou_strategic_2021, simone_museums_2021, leow_analysing_2021, ocon_digitalising_2021, bhaumik_conserving_2020, bozzelli_integrated_2019, marsili_digital_2019, dalbertis_encyclopedic_2018, gonzalez_zarandona_digitally_2018, brumana_virtual_2018, szabo_landscape_2017, vosinakis_visitor_2016, peters_digital_2015, li_interactive_2015, liew_towards_2014, menna_hybrid_2014, dimoulas_audiovisual_2014, smith_participatory_2014, reerink_unicum_2012, styliadis_metadata-based_2009, leonov_laser_2015} \\[6pt]

    \verb|Data Acquisition and Analysis| &~\citep{fontanella_hego_2021, pietroni_interacting_2014, hess_developing_2015, garcia-molina_digitalization_2021, el-hakim_detailed_2004, vilbrandt_digitally_2011, balzani_saving_2024, pistofidis2023design, wu_design_2022, mas_digitization_2022, zhenrao_joint_2021, kuna_interactive_2022, portales_virtual_2021, malik_3d_2021, spagnolo_fringe-projection_2021, de_lazaro_shedding_2021, lassandro_analysing_2021, ocon_digitalising_2021, petrovic_geodetic_2021, zhang_multimedia-based_2021, croix_recasting_2020, ferdani_3d_2020, wijnhoven_digital_2020, loaiza_carvajal_virtual_2020, quattrini_digital_2020, luznik-jancsary_integration_2020, nancarrow_countering_2019, tai_digital_2020, yaagoubi_seh-sdb_2019, rahaman_photo_2019, liang_semantic-based_2019, chng_crowdsourcing_2019, kersten_historic_2018, gil-meliton_historical_2019, adembri_virtual_2018, serain_sensitive_2018, kuroczynski_virtual_2017, rojas-sola_digital_2018, themistocleous_model_2017, liritzis_digital_2015, comes_methodology_2014, georgopoulos_3d_2014, menna_hybrid_2014, styliadis_metadata-based_2009, yastikli_documentation_2007, leonov_laser_2015} \\[6pt]

    \verb|Information Access and Recommendations| &~\citep{li_scene_2023, xu_construction_2023, fan_research_2023, arayaphan2022digitalization, simone_museums_2021, permatasari_web_2020, machidon_culturalerica_2020, serain_sensitive_2018, noardo_architectural_2018, kuroczynski_virtual_2017, cheok_confucius_2017, vosinakis_visitor_2016, giaccardi_social_2008, leonov_laser_2015} \\[6pt]

    \verb|Physical and Virtual Interface Integration| &~\citep{not_digital_2019, montusiewicz_architectural_2021, vilbrandt_digitally_2011, pistofidis2023design, zhenrao_joint_2021, simone_museums_2021, malik_3d_2021, ocon_digitalising_2021, ardito_user-defined_2020, villa_lapidary_2017, brancati_experiencing_2017, liritzis_digital_2015, neamtu_evaluating_2024, krumpen_towards_2021} \\[6pt]

    \verb|Real-time Computing and Multimedia| &~\citep{drygalska_technologically_2023, renzi_storytelling_2023, zhao2023personalized, wu_digital_2023, ocon2023low, quattrini_digital_2020, jaillot_describing_2020, marsili_digital_2019, Massimo_perceiving_nodate, smirnov_context-based_2017, celentano_layered_2017, civantos_using_2016, bartolini_recommending_2016, chen_case_2013, anastasovitis2020creative} \\[6pt]

    \verb|None| &~\citep{konstantakis_adding_2020, jones_enthusiast_2019, li_kano-qfd-based_2023, zhu_designing_2024, aburamadan_developing_2022, cheng_using_2022, zhou_database_2020, xue_evaluation_2019, ricart_promoting_2019, deng_selective_2019, solima_qr_2018, hong_social_2017} \\[6pt]

    \bottomrule
    \end{tabular}
\end{sidewaystable}

In the Sankey Diagram (see Figure~\ref{fig:sankey}), we pay attention to the sample correlation among the three categories so as to conduct a quality assessment of the application types and effects of important links in the digital display of cultural heritage, so as to identify the advantages and disadvantages therein. In this visual classification, we summarize the technologies of different devices in different application fields based on the hierarchical graph model. In it, similar technologies embodied by different devices are aggregated to form an application queue, which facilitates comparison between finding simpler or richer device usage logic in the run-up to the digital presentation or throughout the exhibition process. Specifically, node color represents the refined classification of different devices, applications, and technologies, node height represents the proportion of use of a given device, application, and technology, and the flow line between each pair of nodes represents the proportion of use to the type of the upper level. Based on this, we found that the use of traditional Mobile/PC/laptop terminals is still the largest, and they are widely used in almost all sample areas, especially to enhance the experience of exhibitions and museum~\citep{ze2023prem}. It is worth mentioning that the application of technology in the field of exhibitions and museums is also more concentrated and rich than in other fields. Among them, enhanced and immersive experience technologies contribute the most. In addition, VR devices and AR contribute similarly, but VR devices are more widely used in education and learning. Unfortunately, the application of XR and MR has yet to be expanded, and the use of equipment for material and intangible heritage conservation and management, as well as community and social engagement, is also limited. In addition, we also noticed that the use of Mobile/PC/laptop side of the special population care application has not yet realized the corresponding technical connection.

\subsubsection{Development Trend Analysis}
\begin{figure}[htbp]
    \includegraphics[width=\linewidth]{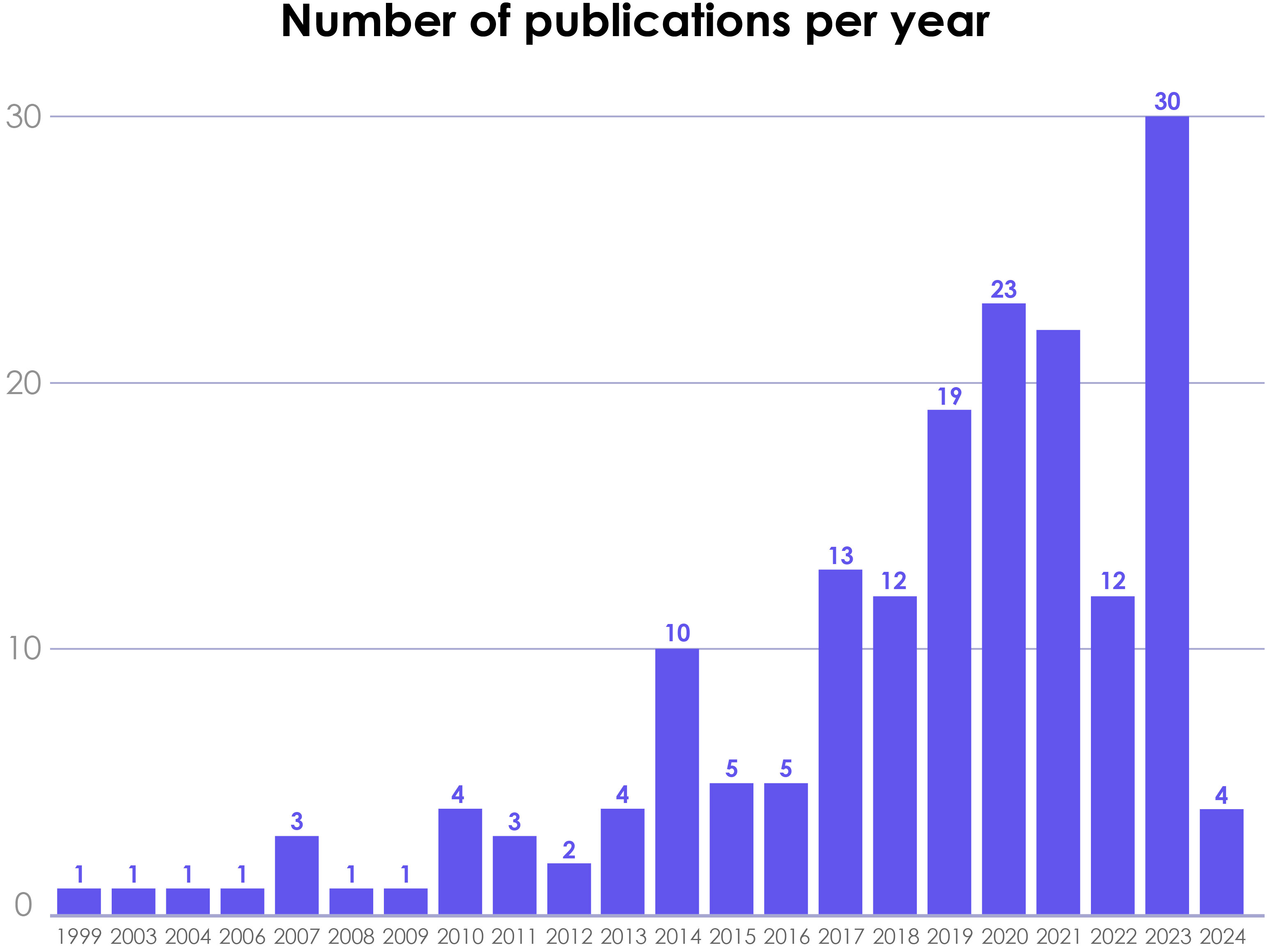}
    \caption{Chart of Trends in the Number of Articles Published in the Field (1999-2024)}
    \label{fig:Development trend analysis}
\end{figure}

In addition, we also counted the publication time of the sample. 
Considering that digitization and immersive presentation in the field of cultural heritage have undergone significant development over time—moving from initial exploratory phases to more mature, diverse applications—we have opted for a longer timeframe to capture this evolution. We aim to cover the early emergence of related research, the iterative technological advancements, the accumulation of practical applications, and the most recent rapid developments in the field. 
We found that the number of publications has been fluctuating at a low value from 1999 to 2013, which may be related to the low level of information technology development at that time. Since 2017, with the rapid improvement of information technology, the number of published articles has also begun to increase significantly, and the blowout growth will be formed in 2023, reaching 30 articles. However, the number of published documents in 2020-2022 has dropped slightly, which we suspect may be related to the impact of the global novel coronavirus epidemic, and the development of relevant technologies is also limited. It is worth noting that the number of 2024 publications shown in the figure is less than 5, because we used the data up to January 2024 when compiling the data, so the 2024 data shown in the figure is not very valuable for reference. However, according to our trend forecast, the publication volume in the relevant field is likely to maintain significant growth in 2024 (see Figure~\ref{fig:Development trend analysis}).

\section{Discussion}

\subsection{Strengths and Benefits of Immersive Cultural Heritage}

Based on the analysis of the sample, we have sorted out and summarized the advantages of the equipment, applications and technologies. It is mainly divided into the following aspects: enlarging the sense of participation and experience, stimulating social attributes, promoting education and knowledge dissemination, optimizing the form of technological realization, and promoting heritage protection and dissemination.
First, through the use of digital technologies and resources, a richer sensory experience can be obtained in an innovative narrative mode in the environment, deepening the memory of visitors for cultural heritage information~\citep{liritzis2015digital}. Personalized, real-time, present and entertaining digital display tools can drive visitors to participate in the exhibition more actively~\citep{buono2022multisensory}.

Secondly, the equipment, technology and application of the digitization of cultural heritage can further stimulate the social character of the exhibition~\citep{bautista2013museums}. For example, in exhibition design, the combination of physical and digital can promote social interaction and community connection, thus promoting cultural transmission and cross-cultural communication, and realizing community co-creation. The use of social tags can make the digital system more accurately capture user interests and enrich user profiles~\citep{oomen2011crowdsourcing}.

Third, in the field of education, digital intervention affects the whole process of knowledge dissemination~\citep{ott2011towards}. In the lead-up period, digital technologies, devices and applications can provide rich learning resources and make them easier to access and discover~\citep{jang_content_2023}. In the process of knowledge transmission, because of the vividness and individuation of the educational experience, the autonomy and efficiency of learning are greatly improved~\citep{semeraro_folksonomy-based_2012}. In addition, in terms of the cultivation of learning thinking, digitization presents a more diverse and fluid knowledge form, helping learners to cultivate a more comprehensive thinking mode, interdisciplinary thinking ability, critical analysis ability, and deep-seated aesthetic experience, thus producing more creative and forward-looking expressions~\citep{kuroczynski_virtual_2017}.

Fourthly, in terms of technology implementation, logical digital methods, clearer visual expression, wide accessibility, research efficiency and accurate implementation reduce the technical threshold to a certain extent, improve the technical acceptance of non-professionals, and provide an easy-to-accept learning tool for the dissemination of cultural heritage~\citep{younes_pose_2016}. On one level, digital technology is cheaper and easier to implement than traditional forms of exhibition.

Fifth, in terms of the protection and dissemination of cultural heritage, digital technology, equipment, and applications create more opportunities for the discovery of details, ensure the integrity and accuracy of data, and enhance the visualization of cultural heritage details to provide more powerful support for the cultural heritage itself~\citep{yastikli_documentation_2007}. In addition, digital tools have opened up new possibilities when visiting cultural heritage, helping to reduce risks, improve accessibility and make it more humane~\citep{liritzis_digital_2015}.

In general, digital technologies and applications have brought unprecedented advantages to the exhibition, dissemination and conservation of cultural heritage. These technologies not only enhance the sensory experience and social interaction of participants but also promote the dissemination of knowledge and the development of education through innovative ways. At the same time, digital technology has lowered the technical threshold of cultural heritage protection, promoted broader audience participation, and provided strong support for the inheritance and development of cultural heritage. Through these technological means, cultural heritage can be more widely disseminated while being protected, enhancing the possibility of cross-cultural communication, and further promoting cultural diversity and global understanding.

\subsection{Challenges and Limitations in Immersive Cultural Heritage}

We have also sorted out and summarized the shortcomings of the equipment, applications and technologies. Mainly divided into the following aspects: technical barriers, sustainability, experience, and safety.
First, technical understanding and operational difficulties may hinder initial users and special groups such as children and the elderly, affecting the access experience. In addition, high digitalization and excessive technological dependence make cultural heritage dependent on specific hardware and software platforms, which can lead to technical failures and compatibility issues~\citep{krol2020digital}.

Secondly, sustainability issues are first reflected in terms of duration, technology and financial investment, which may be applied to the wider field of cultural heritage display~\citep{borin2023financial}. In addition, users are unstable, changing participation, diverse cultural and political differences, and personalized dynamic needs mean that the digital application of cultural heritage needs more long-term and stable support~\citep{borin2023financial}.

Third, the audience's sense of experience will be affected by information overload, and the huge amount of information will make some experience groups feel at a loss. What is more obvious is that the full application of digitalization will lead to the limitation of cultural authenticity and physical experience, and it is difficult to fully convey the cultural background and deep meaning of some traditional crafts, resulting in defects in audience understanding~\citep{corona2023digitization}. Maintaining a balance between virtual content and the real world environment can be a challenge, and too many virtual elements can also distract visitors~\citep{newell2012old}.

Fourth, security issues are reflected in the protection of cultural heritage and copyright issues, user personal data privacy and security issues~\citep{manvzuch2017ethical}. In addition, for some users, the VR experience may cause motion sickness effects and reduce the quality of the experience~\citep{chattha2020motion}.

Overall, while digital technologies bring significant advantages to the display and preservation of cultural heritage, there are still some challenges and limitations that need to be addressed. The complexity and platform dependence of technical operations may limit the participation of some groups, and technical failures and compatibility issues are unavoidable. The sustainability of digital applications is also under long-term pressure from changes in funding, technology and user needs. In addition, information overload, the weakening of cultural authenticity, and the replacement of physical perception by virtual experiences may affect the audience's deep cultural understanding. Security issues can not be ignored, involving the protection of cultural copyright and user privacy~\citep{mulumba2017institutional}. While promoting the digitization of cultural heritage, we must actively address these issues to ensure its long-term stable and healthy development.

\subsection{Gaps and Risks in Current Immersive Cultural Heritage Practices}

We also found some areas that have not been fully explored in the research, and these missing parts may have an important impact on the digital display and dissemination of cultural heritage. It is manifested in the misinterpretation of cultural heritage, the influence of purity in media experience, and the potential damage of cultural heritage.

\subsubsection{Overinterpretation or Misinterpretation of Cultural Heritage in Immersive Representations}

Digital immersion in the presentation of cultural heritage can sometimes lead to over-interpretation or misunderstanding, thus deviating from the original intention of technical intervention~\citep{muir2009eresearch}. For example, one of the purposes of digital artefacts is to deepen the understanding and appreciation of real artefacts. When the focus is no longer on content but on technology, it may not be easy to use and distract visitors from the cultural heritage, leading to technological bias~\citep{havemann2008presentation}.

In addition, 3D reconstruction techniques have been implemented in many significant research projects for digital cultural heritage~\citep{colombo2005metric, keep2022mernda}; however, these techniques sometimes rely excessively on imagination and/or visual additions~\citep{gomes20143d,han2021image}. Spatiotemporal discrepancies, contextual detachment of real data, and technical errors in digital databases can further distort public understanding and compromise authenticity~\citep{colombo2005metric}.

\subsubsection{The Impact on ``Purity in Media Experience''}

The overuse of digital technology may compromise the ``purity in media experience'' in the display of cultural heritage~\citep{laskowska2016purity}. The excessive use of technology may cause the original cultural details of the cultural heritage to be lost and the deep meaning to be tampered with or ignored, resulting in the cultural interpretation and historical background being covered up by the technical display, deviating from the essence of the exhibition. From the audience level, if there are too many digital elements and interactive content, the too rich sensory experience will make it difficult for the audience to concentrate on the exhibition, interfere with the audience's reception of the exhibition content information and in-depth thinking process so that the cultural heritage exhibition will be confined to the technical display level~\citep{havemann2008presentation}.

\subsubsection{Potential Damage to Cultural Heritage from Immersive Practices}

Digital display technologies can also pose risks to cultural heritage. Light and noise pollution, along with excessive visitor interaction, can cause physical damage~\citep{mendez2022attraction}. At the same time, it is also necessary to consider the indirect effects of light on cultural heritage sites, such as the harm caused by the attraction of phototaxis~\citep{verovnik2015reduce}. Inappropriate light can affect the creation of emotions as well as suggest, evoke and support the visitor experience~\citep{di2014innovation}. One example is the night tour at the Longmen Grottoes in Luoyang, China~\citep{liu2012geological}. Field research revealed that the tour was initially suspended after experts warned that the lights attracted insects, which in turn drew animals whose droppings damaged the cave structures~\citep{liu2012geological}. The tour resumed after replacing the light source based on expert advice~\citep{li2024study}. This case shows that the impact of light shows varies depending on the materials of the heritage site, highlighting the need for tailored protection measures~\citep{bista2021lighting}. Similarly, a light show at the Great Wall was halted due to public and expert opposition, for reasons that we do not know, but may have been largely due to concerns that the site was over-commercialized, compromising its conservation, or including excessive brightness, glare, uneven light distribution, inappropriate color temperature, poor color rendering index, as well as resulting in excessive lighting, light clutter, and light intrusion~\citep{zielinska2018historic, bista2021lighting}. These cases illustrate the delicate balance between cultural heritage preservation and public display.

\subsubsection{Other Issues in Immersive Cultural Heritage Practices}

In addition to the previously mentioned risks, several other challenges have surfaced in the application of immersive technologies in cultural heritage preservation and exhibition. These challenges are associated with ethical concerns, economic disparity, and technological standardization, each of which warrants deeper investigation and consideration.

One notable issue is the ethical dilemma associated with the virtual reproduction of sensitive cultural sites. When sacred Spaces or emotionally charged historical events are portrayed in virtual form, people may worry about the trivialization or commodification of cultural experiences. Sites such as war or disaster memorials are critical to the sensitivity of cultural and historical contexts~\citep{arthur2014memory}. The line between educational engagement and the commercialization of tragedy or sacred heritage is often blurred, leading to potential ethical violations~\citep{drakakis2020material}. Another important aspect is that when Western frameworks are used for immersive exhibition and Virtual Representation of indigenous knowledge, this can distort its cultural integrity, hence the need for a more inclusive and participatory approach~\citep{manvzuch2017ethical}.

The rapid adoption of immersive technologies in cultural heritage has also raised concerns about economic inequality~\citep{franks2017desert}. Many cultural institutions, especially those in developing regions, may lack the funds or resources to integrate advanced technologies into their exhibitions~\citep{franks2017desert}. This has created a growing gap between well-funded museums and smaller institutions or sites that rely on traditional methods of preservation and exhibition~\citep{franks2017desert}. In addition, visitors from disadvantaged economic backgrounds may not have access to expensive VR or AR headsets, which may prevent them from having a full range of cultural experiences~\citep{franks2017desert}. Thus, access to the democratization of immersive cultural heritage experiences remains an unsolved challenge~\citep{franks2017desert}.

Another concern is the lack of standardization in the development and application of technologies for immersive experiences. Different institutions often adopt proprietary technologies, leading to compatibility and interoperability issues between digital platforms~\citep{van2024immersive}. This fragmentation can hinder the seamless sharing and preservation of digital cultural heritage across borders, as well as the long-term sustainability of digital archives~\citep{hamad2022virtual}. Without universally accepted standards, organizations may face challenges maintaining or upgrading immersive installations, especially as technology evolves.

While immersive experiences offer new ways to interact with cultural heritage, they also have the potential to disconnect viewers from the material and tactile quality of physical artifacts. The ability to touch, feel, or closely examine the texture and craftsmanship of objects is often necessary for the full appreciation of cultural heritage~\citep{feng2022study}. Immersive technologies can simulate these qualities to a certain extent, but the loss of materiality can distance visitors from the physical reality of heritage, reduce their sensory engagement, and may lead to a superficial understanding of cultural artifacts.

These issues highlight the complexity and risks of using immersive technologies in cultural heritage. While there are many benefits, these emerging challenges must be addressed to ensure that immersive cultural heritage practices remain respectful, inclusive and sustainable.

\section{Conclusion and Future Work}

In this paper we have provided a comprehensive review of the application of immersive technologies in cultural heritage, identifying key opportunities, challenges, and gaps in the field. Analyzing over 5,000 articles, we distilled insights from 177 key papers to map the current landscape of immersive technologies, including VR, AR, and MR, and their use in cultural heritage. Our findings highlighted the transformative impact of these technologies in enhancing preservation, education, and cultural dissemination, offering personalized, interactive experiences that deepen audience engagement and promote cross-cultural dialogue.

However, grand challenges remained, including technical barriers such as platform dependence and interoperability, sustainability concerns, and risks of diminishing cultural authenticity through digital reconstructions. \textbf{Data privacy} and \textbf{misrepresentation of cultural heritage} are critical considerations that must be addressed alongside other ethical issues. Ensuring responsible use of these technologies requires a balanced approach that respects both the technical and ethical dimensions. Furthermore, addressing these areas necessitates ongoing research to ensure that the development of digital and immersive display technologies honors the essence of cultural heritage and aligns with societal expectations.

Future research should focus on addressing these challenges through sustainable, scalable solutions, balancing innovation with cultural authenticity, and fostering collaborations between cultural heritage experts and technologists. Additionally, greater awareness of the ethical and social implications of these technologies is needed, especially in regard to cultural representation and intangible heritage. By overcoming these challenges, immersive technologies can reshape how we engage with cultural heritage, creating more inclusive and globally accessible experiences. Our findings offer a roadmap for future research and practice in this evolving field. 

We also plan to organize a workshop with practitioners in cultural heritage, as well as experts in VR/AR and immersive interactive technologies, to exchange ideas on creating more inclusive and engaging experiences for users. We aim to collaborate with governments and relevant institutions (for example, museums) to develop more comprehensive cultural heritage experiences, including advanced equipment and dedicated venues.

\backmatter

\phantomsection
\addcontentsline{toc}{section}{Data availability}
\bmhead{Data availability}
All literature analyzed during this project is at \url{https://www.repository.cam.ac.uk/handle/1810/379779}~\citep{wang_2025}.

\phantomsection
\addcontentsline{toc}{section}{Acknowledgements}
\bmhead{Acknowledgements}

This work was supported in part by a grant from Research Project of the National Museum: `Research on Intelligent Interactive Design for Museum Exhibitions Based on Audience Experience' (\# GBKX2021Z10).

\phantomsection
\addcontentsline{toc}{section}{Author Information}
\bmhead{Author Information}

\paragraph{Hanbing Wang} is a Research Assistant at Tsinghua University. She received her BA degree from Beihang University, P.R. China, and her MA degree from Goldsmiths, University of London, U.K. Her research interests include digital innovation and technology in cultural heritage, immersive experience design in new media, and interaction design theory and methodology.

\paragraph{Junyan Du} is a Master's student at Tsinghua University. She obtained a Bachelor's degree in Visual Communication Design from China University of Geosciences (Wuhan).

\paragraph{Yue Li} is an Assistant Professor in the Department of Computing at Xi’an Jiaotong-Liverpool University. Her research interest is in the field of human-computer interaction, with particular emphasis on the design, evaluation, and application of virtual and augmented reality technologies in cultural heritage and education.

\paragraph{Lie Zhang} is a Professor at Tsinghua University. He received his Bachelor's degree from Zhejiang University and his Master's and Doctoral degrees from Tsinghua University. Since 2000, he has been teaching at Tsinghua University. His research interests include information and interaction design, media space art, cultural heritage digital display, and autism rehabilitation.

\paragraph{Xiang Li} is a PhD student in the Department of Engineering at the University of Cambridge and a Student Fellow at The Leverhulme Centre for the Future of Intelligence. He received his BSc (Hons) in Information and Computing Science from Xi'an Jiaotong-Liverpool University and the University of Liverpool.

\bibliography{sn-bibliography} 

\end{document}